\documentclass[aps,prl,twocolumn,showpacs,floatfix,a4paper]{revtex4}
\usepackage{graphicx}
\usepackage{amsmath}
\usepackage{amssymb}
\usepackage{bm}
\usepackage[usenames]{color}
\usepackage{amsfonts}
\usepackage{appendix}
\usepackage{xfrac}
\usepackage[normalem]{ulem}

\graphicspath{ {images/} }
\let\oldsqrt\sqrt
\def\sqrt{\mathpalette\DHLhksqrt}
\def\DHLhksqrt#1#2{\setbox0=\hbox{$#1\oldsqrt{#2\,}$}\dimen0=\ht0
\advance\dimen0-0.2\ht0
\setbox2=\hbox{\vrule height\ht0 depth -\dimen0}{\box0\lower0.4pt\box2}}
\DeclareMathSizes{10}{10}{6}{5}
\newcommand{\rpm}{\raisebox{.2ex}{$\ \scriptstyle\pm$}}
\newcommand{\rpp}{\raisebox{.2ex}{$\scriptstyle +$}}
\newcommand{\rmm}{\raisebox{.2ex}{$\scriptstyle -$}}

\begin{document}

\title{Transmission Resonances Anomaly in 1D Disordered Quantum Systems}
\author{A. Eisenbach$^{1}$}
\author{Y. Bliokh$^{2}$}
\author{V. Freilkher$^{1}$}
\author{M. Kaveh$^{1}$}
\author{R. Berkovits$^{1}$}
\affiliation{$^1$Department of Physics, Jack and Pearl Resnick Institute, Bar-Ilan University, Ramat-Gan 52900, Israel\\
$^2$Department of Physics, Technion-Israel Institute of Technology, Haifa
32000, Israel}

\begin{abstract}
Abstract
\end{abstract}

\begin{abstract}
Connections between the electronic eigenstates and conductivity of
one-dimensional disordered systems is studied in the
framework of the tight-binding model. We show that for weak disorder
only part of the states exhibit resonant
transmission and contribute to the conductivity. The rest of the eigenvalues are not associated with peaks in transmission and the amplitudes of their wave functions do not exhibit a significant maxima
within the sample. Moreover, unlike ordinary states, the lifetimes of
these `hidden' modes either remain constant or even decrease (depending on
the coupling with the leads) as the disorder becomes stronger. In a wide
range of the disorder strengths, the averaged ratio of the number of
transmission peaks to the total number of the eigenstates is independent of
the degree of disorder and is close to the value $\sqrt{2/5}$, which was derived
analytically in the weak-scattering approximation. These results are in
perfect analogy to the spectral and transport properties of light in
one-dimensional randomly inhomogeneous media \cite{Bliokh2015}, which
provides strong grounds to believe that the existence of hidden,
non-conducting{\large \ }modes is a general phenomenon inherent to 1D open
random systems, and their fraction of the total density of states is the
same for quantum particles and classical waves.
\end{abstract}

\maketitle

\section{\protect\bigskip Introduction}

In a recent paper \cite{Bliokh2015}, an interesting find regarding the
transmission of waves through disordered systems has been presented. It has
been shown analytically, numerically and experimentally that in weakly
disordered one-dimensional dielectric media, a substantial fraction of
optical quasi-normal modes (QNMs) are hidden, i.e., could not be detected by
transmission measurements. Such a behavior should be expected also for the
transmission of other waves, particularly for the electron transport in
disordered conductors. States of open electronic system can also be
interpreted in terms of QNMs \cite{Pnini1996,Ching1998,Leung1994}, which from the
mathematical point of view are the generalization of the notion of the
eigenstates of closed (Hermitian) quantum-mechanical structures. The
imaginary parts of the eigenvalues of a non-Hermitian Hamiltonian depict the
lifetimes of the QNMs \cite{Wang2011,Wang2012}, which are finite due to the flow of
electrons between leads. Therefore, recasting the classical problem
considered in \cite{Bliokh2015} for electronic systems is of interest, since
one can ask additional questions regarding QNMs, which are difficult or
non-relevant in optics. Especially, one can probe the hidden modes (HM)
response to non-equilibrium conditions such as a large applied source-drain
voltage, temperature, interaction with other electrons or phonons, etc. Here
we study the electronic spectra of one-dimensional disordered systems in the
non-equilibrium Green function (NEGF) formulation, which enables us to address
the problems unique to electronic transmission.

In open homogeneous structures like clean quantum wires, open
resonators, etc., to each QNM corresponds a transmission resonance (TR)
(peak in the frequency spectrum of the transmission coefficient) with the
resonant energy equal to the real part of the eigenvalue \cite{Moiseev2011}.
This is not necessarily the case in open \textit{disordered} samples. In the
presence of disorder the position and height of the TR fluctuate,
a phenomena associated with mesoscopic conductance
fluctuations \cite{Beenaker1997,Imry1999}. Here we show that one-to-one correspondence
between the number of QNMs and TRs could be broken as well. Due to complex interference between multiply scattered random fields, in weakly disordered systems some of QNMs become
invisible in transmission (hidden), and the number of the transmission peaks
falls to $\sqrt{{2}/{5}} \cdot N$ (where $N$ is the total number of QNMs).

Although there is a common believe that after more than fifty years of
intensive study the transport
properties of 1D disordered systems are clearly understood, surprisingly
enough, the existence of the hidden modes was completely overlooked. This is
perhaps because the attention was mostly concentrated on the localization at
strong disorder, while the limit of weak impurities (ballistic regime) was
deemed trivial.

In the present paper, we investigate the evolution of the transmission and
of the density of states (DOS) of quantum-mechanical particles in a random 1D
potential (tight-binding wire), for a wide range of the disorder strengths,
from ballistic to strong localization regimes. We show that the coexistence
of two types of QNMs (ordinary and hidden) is rather general phenomenon
intrinsic to randomly inhomogeneous one-dimensional quantum-mechanical
systems as well. Not only do the hidden electron states exist and manifest
analogues properties as the corresponding solutions of Maxwell equations,
the relative number of hidden states for weak and moderate disorder is also the same. Its mean value in a given energy interval remains close to the constant $1 - 
\sqrt{{2}/{5}}$ over wide ranges of disorder strengths and of the
length of the system. The value $1 - \sqrt{2/5}$ follows from
general statistical properties of random trigonometric polynomials.

Furthermore, in contrast to the well-known behavior of the localized states, the
lifetime of a hidden state does not rise with increasing fluctuations of the
potential, but rather remains unchanged or even decreases, depending on the
strength of the coupling to the leads. The eigenvectors (solutions of the
Schrodinger equation satisfying the outgoing boundary conditions) of such
modes are also very unusual. The spatial profiles of their
amplitudes are neither concentrated near both edges of the system with a
minimum in the center as in symmetric clean systems,
nor are they localized as in a potential with strong fluctuations. On the
contrary, the wave functions of the hidden states nestle up near one of the
edges of the wire and exponentially decreases towards the other.

As the scattering strength and the length of the system increase, hidden
modes eventually become ordinary. An important feature of HMs, specific for
electronic systems is that although they appear in the DOS in the same way
as the ordinary modes do, they are non-conducting, i.e. do
not contribute to the conductivity even in the ballistic regime. The quantum
mechanical treatment of these hidden QNM by the NEGF method enables a
simple analysis of their spatial behavior. We show that the TR anomaly is
directly related to hybridization with the leads, and therefore it
becomes more subtle at higher disorder and vanish where the localization length is shorter
than the system length.

In the next two sections, we introduce the model and overview the NEGF
method. In the fourth section, we show the lateral behavior of the hidden
QNM, the counter intuitive dependence on the strength of disorder, and the
impact of temperature on the TR counting. In the appendix an analytical derivation of the ratio $N_{T\!R}/N$ in the single-scattering approximation is presented.

\section{The model}

Here we consider a one-dimensional (1D) wire, coupled to two semi-infinite
leads on the left and on the right. The disordered tight-binding Hamiltonian
of the wire is given by \cite{Anderson1961}:

\begin{equation}  \label{Eq:HamiltonianWire}
\hat{H}_{w}=\displaystyle\sum_{j=1}^{L}\varepsilon _{j}{\hat{c}}%
_{j}^{\dagger }{\hat{c}}_{j}-\left( t\displaystyle\sum_{j=1}^{L-1}{\hat{c}}%
_{j}^{\dagger }{\hat{c}}_{j+1}+h.c.,\right) .
\end{equation}%
where ${\hat{c}}_{j}$ is the single-particle annihilation operator on site $%
j $; $t$ is the hopping amplitude, which is set to 1 throughout the paper. The on-site potentials $\varepsilon _{j}$%
\ are statistically independent random numbers homogeneously distributed in
the range $[-W/2,W/2]$. 
As long as the wire is not connected to the leads, $\hat{H}_w$ can be numerically diagonalized and its eigenvalues $E_i$ and eigenvectors $\psi_i(j)$ may calculated.

The left and right leads are represented by the Hamiltonians: 
\begin{equation}
\hat{H}_{l/r}=-t\displaystyle\sum_{j=1}^{\infty }{\hat{c}_{j}^{(l/r)\dagger }%
}{\hat{c}_{j+1}^{(l/r)}}+h.c.,
\end{equation}%
where ${\hat{c}_{j}^{(l/r)}}$ is the\emph{\ }single-particle annihilation
operator on site $j$ of the left ($l$) or right ($r$) lead, $t$ is the same
hopping amplitude as in the wire, and there is no on-site potential in the
leads. The left/right lead is coupled to the wire by: 
\begin{equation}
\hat{H}_{w,l/r}=-t_{l/r}{\hat{c}_{1}^{(l/r)\dagger }}{\hat{c}_{(1/L)}}+h.c.,
\end{equation}%
where $t_{l/r}$ is the coupling\emph{\ }amplitudes\emph{\ }between the
left/right lead and the wire. Thus, the complete Hamiltonian of the system
composed of the wire and leads is given by: 
\begin{equation}  \label{Eq:Hamiltonian}
\hat{H}=\hat{H}_{w}+\hat{H}_{l}+\hat{H}_{r}+\hat{H}_{w,l}+\hat{H}_{w,r}.
\end{equation}

\section{\protect\bigskip}

\section{Transmission function and the density of states}

The quantities of interest, namely the transmission function of the wire $T_{lr}$ \ and the density of states $\mathcal{N}(E)$, can be expressed through the tensor Green's function $G$, whose $G_{ij}$ component represents
the probability of a particle to propagate from site $i$\ to site $j$. as follows: 
\begin{equation}
T_{ij}\propto \left\vert G_{ij}\right\vert ^{2}  \label{Eq:Transmission1}
\end{equation}
\begin{equation}
\mathcal{N}(E)\propto \mathrm{Tr}\left( \mathrm{Im}G\right)
\label{Eq:Spectral1}
\end{equation}

Therefore we first calculate the Green's function of the infinite wire-leads
system using the NEGF method. In the following derivation we follow the path and notations presented in Ref. \cite{Datta1995}.

First we present the general form of the Green's function
\begin{equation}
\hat{G}=\left[ E\hat{I}-\hat{H} \rpm\ i\eta\hat{I} \right] ^{-1}  
\label{Eq:GA}
\end{equation}
where $\eta $ is an infinitesimal positive number, $\hat{I}$ is the identity
matrix and $\hat{H}$ is the Hamiltonian (Eq. (\ref{Eq:Hamiltonian})). $\rpp i\eta$ is associated with the retarded Green's functions ($\hat{G}^R$) and $\rmm i\eta$ with the advanced Green's functions ($\hat{G}^A$).
Obviously, directly solving the Green's function requires the inversion of
the infinite matrix $\left[ E\hat{I}-\hat{H} \rpm\ i\hat{I}\eta \right] $.

To proceed, we express $\hat{G}$ through the Green's functions of its
components, i.e., the wire ($\hat{G}_{w}$) and the left ($\hat{G}_{l}$) and
right ($\hat{G}_{r}$) leads. These Green's functions can be written in the following form: 

\begin{multline}
\hat{G}=
\begin{pmatrix}
\hat{G}_{l/r} & \hat{G}_{l/r,w} \\ 
\hat{G}_{w,l/r} & \hat{G}_{w}%
\end{pmatrix}%
= \\
\begin{pmatrix}
\ \left[ \left( E \rpm\ i\eta \right) \hat{I}-\hat{H}_{l/r}\right] & \hat{\tau}
_{l/r} \\ 
\hat{\tau} _{l/r}^{\dagger } & \left[ E\hat{I}-\hat{H}_{w}\right]%
\end{pmatrix}%
^{-1},  \label{Eq:GR_sys_lead}
\end{multline}
where the matrices $\hat{\tau} _{l/r}$ have a single non-zero element $\hat{\tau}_{l}(1,1)=\hat{\tau}_{l}^{\dagger}(1,1)=t_{l}$ and $\hat{\tau}_{r}(L,L)=\hat{\tau}_{r}^{\dagger}(L,L)=t_{r}$. Multiplying
both sides by the inverse right-hand matrix results in two independent
equations for $\hat{G}_{w}$: 
\begin{equation}
{\hat{\tau} _{l/r}}^{\dagger }\hat{G}_{l/r,w}+\left[ E\hat{I}-\hat{H}_{w}%
\right] \hat{G}_{w}=\hat{I},  \label{Eq:GR_sys_lead2}
\end{equation}%
\begin{equation}
\left[ \left( E \rpm\ i\eta \right) \hat{I}-\hat{H}_{l/r}\right] \hat{G}%
_{l/r,w}+\hat{\tau}_{l/r}\hat{G}_{w}=0.  \label{Eq:GR_sys_lead3}
\end{equation}%
Combining the two equations and taking into account both leads one gets 
\begin{equation}
\hat{G}_{w}=\left[ E\hat{I}-\hat{H}_{w}-\hat{\Sigma}\right] ^{-1}
\label{Eq:GR_sys}
\end{equation}%
where the total self-energy equals $\hat{\Sigma}=\hat{\Sigma} _{l}+\hat{\Sigma}_{r}$ 
and $\hat{\Sigma}_{l/r}$ is given by 
\begin{equation}
\hat{\Sigma} _{l/r}=\hat{\tau} _{l/r}^{\dagger }\left[ \left( E \rpm\ i\eta \right) 
\hat{I}-\hat{H}_{l/r}\right] ^{-1}{\hat{\tau}_{l/r}},  \label{Eq:GR_Sigma}
\end{equation}%
while $\rpp i\eta$ and $\rmm i\eta$ refer to $\hat{\Sigma}^R_{l/r}$ and $\hat{\Sigma}^A_{l/r}$, respectively.

Since for a 1D lead $\hat{\tau}_{l}$ has only one diagonal non-zero term, the
relevant element in the left lead Green's function $\hat{G}_{l}$ is
the $(1,1)$ element. For a semi-infinite lead it can be calculated
analytically, 
\begin{equation}
\left[ \left( E \rpm\ i\eta \right) \hat{I}-\hat{H}_{l}\right] ^{-1}(1,1)=-\frac{1%
}{t}e^{\pm ika},  \label{Eq:GR_lead4}
\end{equation}%
where $a$ is the lattice constant and $k$ is the wave number of the
electron, which obeys tight-binding dispersion relation $E=-2t\cos (ka)$.
Therefore, the self-energy has also a single non-zero term: 
\begin{equation}
\hat{\Sigma} _{l}(1,1)=t_{l}^{2}\left( -\frac{1}{t}e^{\pm ika}\right)
\label{Eq:GR_Sigma1}
\end{equation}

In the same way, the single non-zero term of the right lead self-energy $%
\Sigma _{r} $ is equal to

\begin{equation*}
\hat{\Sigma}_{r}(L,L)=t_{r}^{2}\left( -\frac{1}{t}e^{\pm ika}\right) .
\end{equation*}

It can be shown \cite{Datta1995} that the transmission through the wire is equal to\bigskip 
\begin{equation}
T_{lr}=\mathrm{Tr}\left[ \hat{\Gamma} _{l}\hat{G}_w^{R}\hat{\Gamma} _{r}\hat{G}_w%
^{A}\right]  \label{Eq:Transmission2}
\end{equation}%
where 
\begin{equation}
\hat{\Gamma} _{l/r}=i\left[ \hat{\Sigma} _{l/r}^{R}-\hat{\Sigma} _{l/r}^{A}%
\right]= -2\cdot \mathrm{Im}\left(\hat{\Sigma} _{l/r}^{R}\right) ,
\label{Eq:GR_Gamma}
\end{equation}%
which results in 
\begin{equation}
T_{lr}=\left( \frac{t_{l}t_{r}}{t}\right) ^{2}\cdot \left\vert \hat{G}_w%
^{R}(1,L)\right\vert ^{2}.  \label{Eq:Transmission3}
\end{equation}
\bigskip

For the calculation of the total current through the system, the population
in the leads and the applied voltage should be taken into account. Assuming
that the leads are in thermal equilibrium at temperature $\mathcal{T}$, the
probabilities to find an electron at a state with an energy $E$ in the left
(right) lead is given by the Fermi distributions{\LARGE \ }$f_{l}$ \ ($f_{r}$%
), and depends also on the electro-chemical potential in the leads $\mu$ ($%
\mu -V$), where $V$ is the voltage drop between the leads. Here we assume that
there are no incoherent effects, such as electron-electron or
electron-phonon interactions, and therefore the total coherent current is: 
\begin{equation}
I=\int {\frac{2e^{2}}{h}T_{lr}(E)\left[ f_{l}\left( E,\mu ,\mathcal{T}%
\right) -f_{r}\left( E,\mu -V,\mathcal{T}\right) \right] dE}.
\label{Eq:Current}
\end{equation}

Consequently, at zero temperature the electrical conductance $g=\partial
I/\partial V|_{V\rightarrow 0}$ is proportional to the transmission
function, $g\propto T_{lr}(\mu )$ . Yet, at finite temperatures and given
source-drain voltages, Eq. (\ref{Eq:Current}) leads to a non-trivial relation,
and numerically one usually performs a finite derivative of the current in
order to calculate the conductance.

In a \textit{clean} wire ($W=0$) with perfect coupling to the leads $%
t_{l/r}=t$, the transmission equals $1$ for all energies in the band $-2t<E<2t.$ At lower coupling \ to leads{\LARGE \ } $t_{l}=t_{r}<t$, the
transmission, as well as the conductance, are non-monotonic functions of
energy, peaked at the eigen-energies of the disconnected wire. The number of peaks is equal to the number of states in the disconnected wire, which are the electronic equivalent of the normal modes of a closed optical cavity.

The local density of states (LDOS) for an isolated wire whose Hamiltonian $%
H_{w}$ is given by Eq. (\ref{Eq:HamiltonianWire}), is equal to 
\begin{equation}
\rho \left(j,E\right) =\sum_{i}\left\vert \psi _{i}(j)\right\vert ^{2}\delta
(E-E_{i}).  \label{Eq:DOS}
\end{equation}

Once the wire is connected to the leads the delta function broadens and the
LDOS is expressed via the spectral function defined as: 
\begin{equation}
\hat{A}=i\left[ \hat{G}_w^{R}-\hat{G}_w^{A}\right] = -2\cdot \mathrm{Im}\hat{G}_w%
^{R},  \label{Eq:Spectral2}
\end{equation}%
so that the diagonal element $\hat{A}(j,j)$ represents the LDOS $\rho \left(
j,E\right) $, while its trace is the DOS
\begin{equation}
\mathcal{N}(E)=\frac{1}{2\pi }\mathrm{Tr}\hat{A}  \label{Eq:Spectral3}
\end{equation}%
.

\section{Transmission resonances}

In an isolated wire composed of $L$ sites with random potentials, the eigenstates vary with the on-site disorder strength, yet each state has a real energy eigen-value, and the DOS $\mathcal{N}(E)$ follows Eq.(\ref{Eq:DOS}). 

Once the wire is coupled to the leads the eigenvalues are complex and states may overlap, nevertheless the DOS can be defined (see Eq. (\ref{Eq:Spectral3})). $\mathcal{N}(E)$ shows peaks at energies close to the eigenvalues $E_i$, with broadening which become wider as $t_{l/r}$ approaches $1$. The total number of states (quasi-normal modes) is given by the integration $N_{Q\!N\!M}$ $=\int_{-\infty }^{\infty}{\mathcal{N}(E^{\prime })dE^{\prime }}$. Obviously, the conservation of degrees of freedom oblige  $N_{Q\!N\!M}=L$.

Similarly, the transmission function $T_{lr}$ in the open and disordered system shows sharp resonances located close to the eigen-energies of the wire $E_i$, with exponentially low valleys between them.
Naturally, the mean value of the transmission is attenuated as the disorder increased and can be scaled by $T_{lr}\sim\exp(-\L/\xi)$, where $\xi$ is the localization length (in the 1D case $\xi \approx 10^2/W^2$ \cite{Romer1997}).

However, in contrast to the DOS, the transmission significantly changes for an open wire, as some of the peaks which existed at the clean wire disappear.


\begin{figure}[tbp]
\centering
\includegraphics[width=8.5cm,height=!]{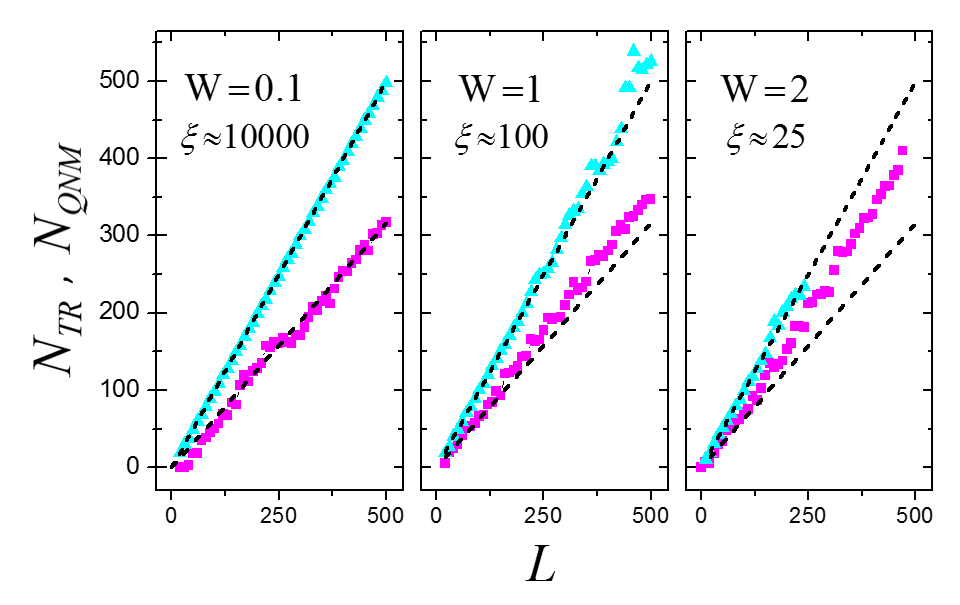}
\caption{The number of the transmission maxima $N_{T\!R}$ (magenta squares) and the
number of quasi normal modes, $N_{Q\!N\!M}$, calculated by integration over the density of states (cyan triangles) for a disordered 1D wire as a
function of the length $L$. The cases of: low disorder $\protect\xi \approx
10000$ (left); medium disorder $\protect\xi \approx 100$ (middle); and
strong disorder $\protect\xi \approx 25$ (right) are presented. In the low
disorder case the number of transmission peaks fits to $N_{\mathrm{tr}}=%
\protect\sqrt{2/5} L$ (lower black dashed lines), while the integrated
density of states follows $L $ (upper black dashed lines). At higher
disorders more transmission resonances are seen (i.e., $N_{\mathrm{tr}}>\protect\sqrt{2/5} L$)
due to localization.}
\label{Fig1}
\end{figure}

In Fig. \ref{Fig1} we present the results for the number of QNMs, $N_{Q\!N\!M}$, and for the number of the transmission resonances (maxima in $T_{lr}\left( E\right) $), $N_{T\!R}$, as functions of the wire size $L$ for different strengths of disorder (Here and in the remainder of the paper all lengths are presented in units of the lattice constant $a$ which is set to unity).

As can be seen, 
the dependence of the number of the transmission resonances, $N_{T\!R}$, on $L$ 
is quite different from that
of the number of QNMs. For weak disorder ($W=0.1$, localization length $\xi
\sim 10^{4}\gg L$) , $N_{T\!R}$ is much smaller than $N_{Q\!N\!M}$ and equals to $\sqrt{2/5}L$.
The rest of the QNMs are hidden, exactly as it is in optical systems
considered in Ref. \cite{Bliokh2015}. As the disorder becomes stronger, the
hidden (with no transmission resonances) modes gradually reappear as peaks in the transmission function. 
This can be seen in the increase of the slope of $N_{T\!R}$ versus $L$ dependence with
increasing $W$. For stronger disorder this ratio tends to one.

\begin{figure}[tbp]
\centering
\includegraphics[width=8.5cm,height=!] {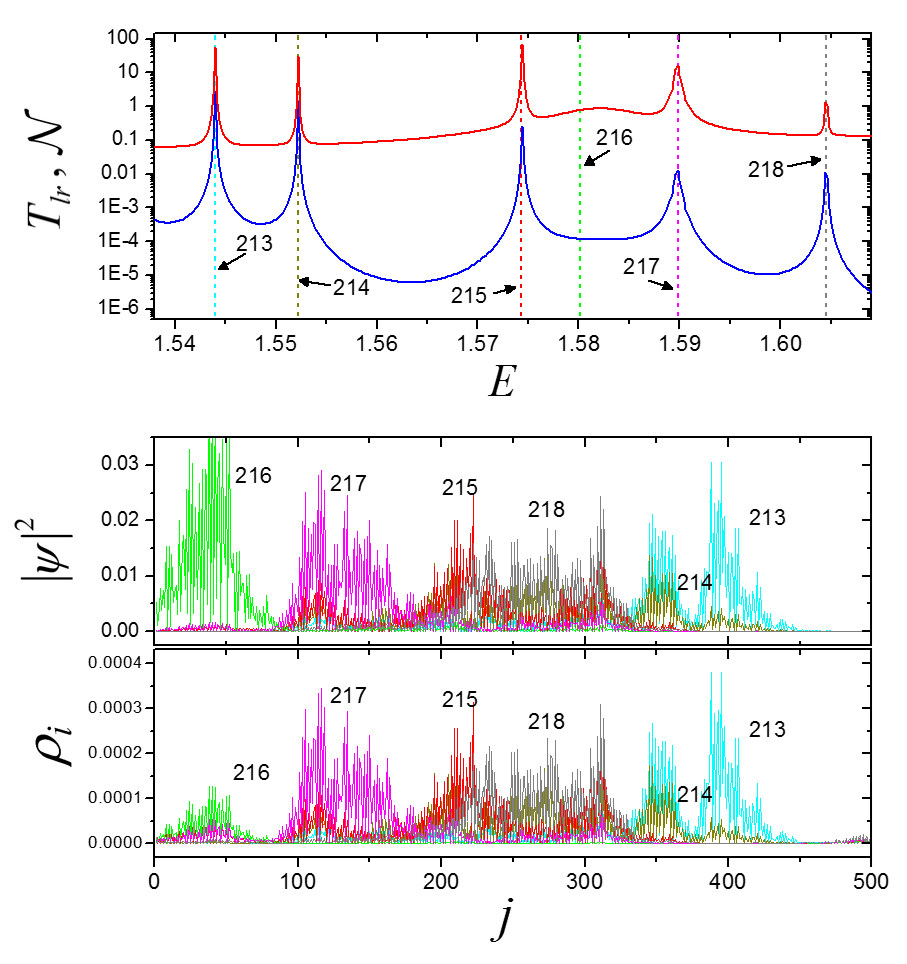}
\caption{\textbf{Upper frame:} Typical density of states $\mathcal{N}(E)$ (red
solid line) and transmission $T_{lr}\left(E\right)$ (blue solid line)
spectra of a particular realization of disorder ($L=500$, $W=1$%
, $t_{l/r}=1$). The positions of the isolated Hamiltonian eigenvalues $%
\protect\epsilon_i$ are indicated by the vertical dashed lines. \textbf{Middle frame:} The squared eigenvectors $|\protect\psi_i(r)|^2$ of the isolated
Hamiltonian as function of the position along the wire, $j$. The 216-th eigenstate is located close to the system edge and
therefore its transmission resonance is washed out (see upper frame) when
the wire is coupled to the leads. \textbf{Lower frame:} The local density of states
integrated in the vicinity of the $i$-th disconnected eigenvalue $\protect%
\epsilon_i$, $\protect\rho_i (r) = \protect\int_{\protect\epsilon%
_i-\Delta/4}^{\protect\epsilon_i+\Delta/4} \protect\rho \left( r, E \right)
dE$. For most states $\protect\rho_i (r) \sim |\protect\psi_i(r)|^2$, except
for the hidden mode (the 216-th eigenstate) for which the local density
close to the leads is strongly suppressed.}
\label{Fig2}
\end{figure}

\begin{figure}[tbp]
\centering
\includegraphics[width=8.5cm,height=!] {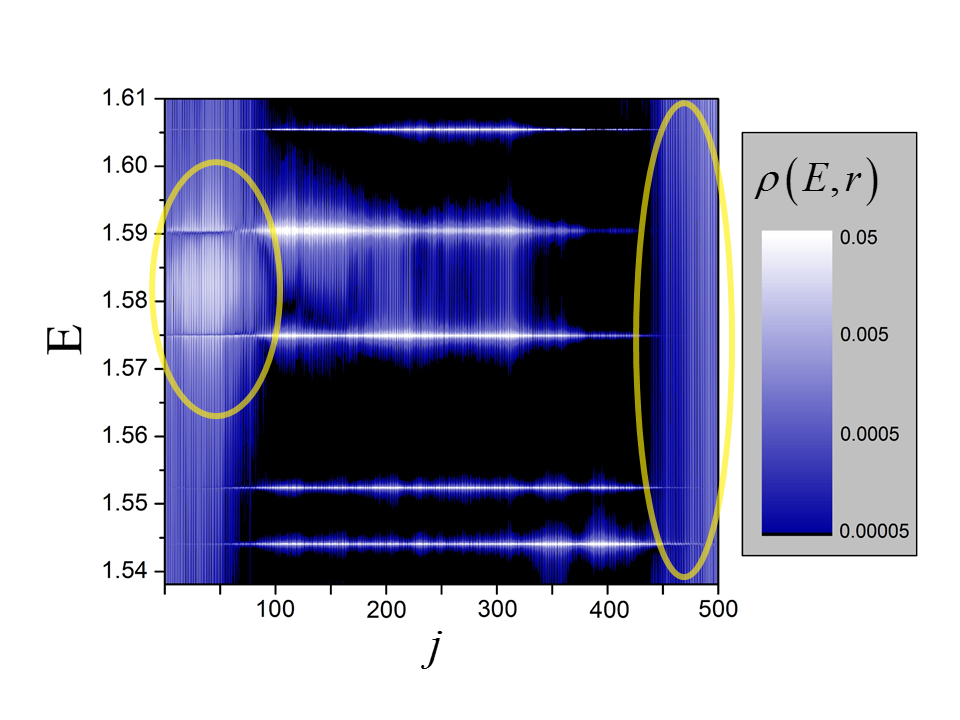}
\caption{A color map of the local density of states $\rho(j,E)$ of the system
described in Fig. \protect\ref{Fig2}. The `ordinary' modes (at $E \sim
1.544, 1.552, 1.575, 1.59, 1.604$) show relatively narrow energy
distribution, while the hidden mode originally located at $E \sim 1.58$ (marked with yellow circle on the left) 
is significantly broadened due to the coupling to the left lead. Similar
hidden mode's tail can be noticed at the right end, related to a state hidden at higher energy (long yellow circle).}
\label{Fig3}
\end{figure}

To understand the nature of the `hidden' states let
us juxtapose the transmission peaks of the eigen-vectors of the disconnected
wire. In the upper panel in Fig. \ref{Fig2} we plot $\mathcal{N}(E) $ and $%
T_{lr}\left( E\right) $ for a typical realization of disorder in a $L=500$
wire with $W=1$, $\xi \sim 10^{2}$. The corresponding modulus-squared
eigen-vectors for the isolated system $|\psi _{i}(r)|^{2}$ are plotted in
the middle panel. It is easy to see that each transmission peak (and the
associated peak in DOS) corresponds to an eigenstate of the isolated wire,
and the peaks in $\mathcal{N}(E)$ and $T_{lr}(E)$ 
are close to the real eigenvalue $\epsilon _{i}$
(indicated by vertical dashed lines). However the hidden state
\#216 does not show any peak in the transmission, and the DOS exhibits
only a very broad maxima at this eigenvalue.
The distinction between hidden and ordinary states shows up also in the local
density of states, which for an $i_{th}$ eigenstate we define 
as $\rho_{i}(r)=\int_{\epsilon _{i}-\Delta /4}^{\epsilon _{i}+\Delta /4}\rho \left(r,E\right) dE$ ,where $\epsilon_i$ is the level's-eigen energy and $\Delta $
is the level spacing.
Indeed, while for the ordinary states the
local DOS of the connected wire is similar to the density of the
disconnected wire i.e., $\rho _{i}(r)\sim |\psi _{i}(r)|^{2}$, for the
hidden mode (state $216$) there is a huge difference between $\rho
_{216}(r)$ and $|\psi _{216}(r)|^{2}$ (see lower frame of Fig. \ref%
{Fig2}).

In Fig. \ref{Fig3}, the LDOS map of the above system in the relevant
energy range is presented. The hidden mode originally located at $E=1.58$
(\#216) is broadened much beyond the mean level spacing. The spatial
distributions of the two types of states are also quite different. Namely,
the hidden ones are always nestled against an edge of the sample, so
that when the wire is coupled to the leads, these modes become strongly
hybridized with the states of the neighboring lead and do not reach the
opposite edge of the sample.

\bigskip 
\begin{figure}[tbp]
\centering 
\includegraphics [width=8.5cm,height=!]{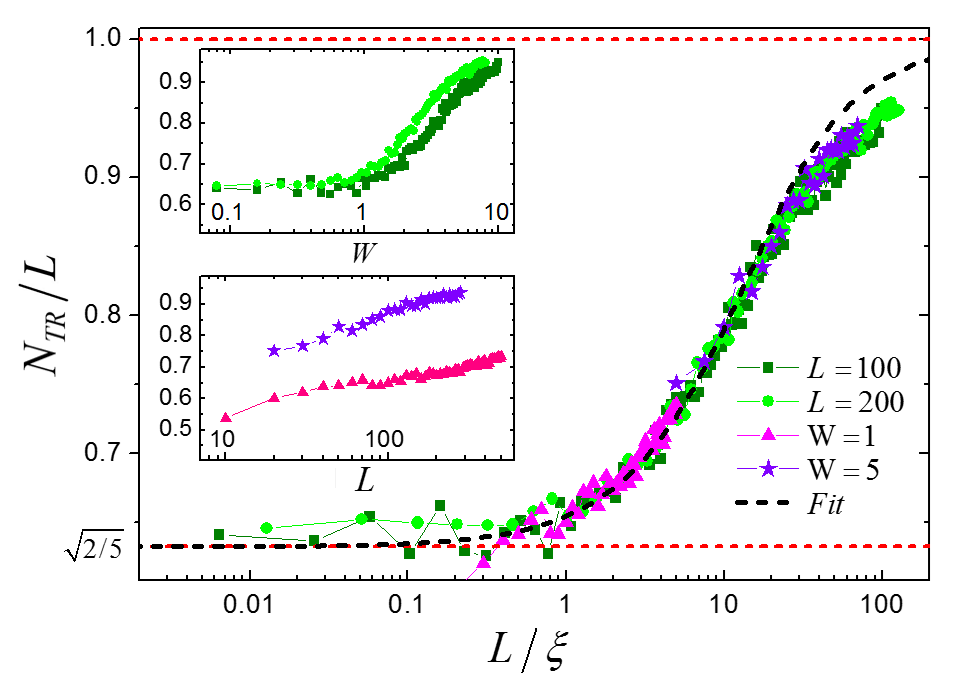}
\caption{ The ratio of the number of observed transmission peaks to the
length of the wire $N_{\mathrm{tr}}/L$ for various disorder strength, $W$, and
wire length $L$. \textbf{Upper inset:} Systems with length $L=100$ and $L=200$ for various disorder 
values. \textbf{Lower inset:} Systems with
disorder strength $W=1$ and $W=5$ for various lengths.
\textbf{Main panel:} the ratio $N_{\mathrm{tr}}/L$ as a function of the scaling parameter $L/\protect\xi$ for the results presented in the insets. All curves of $N_{\mathrm{tr}}/L$ fall
on top each other. For $L/\protect\xi <1 $, $N_{\mathrm{tr}}/L\sim \protect%
\sqrt{2/5}$ remains. Once $L/\protect\xi >1$ , the ratio increases until $%
N_{\mathrm{tr}}/L\rightarrow 1$ for large values of $L/\protect\xi $, i.e.,
for strong localization all modes have transmission resonances. The black
dashed line represents the dependence of the $N_{\mathrm{tr}}/L$ on $L/%
\protect\xi $ according to Eq. (\protect\ref{Eq:OverlapCount}-\ref{Eq:FinalFit}) with $b=1/4$.
}
\label{Fig4}
\end{figure}

Numerical calculations show that at weak disorder, when where $\xi$ is
larger than the system size, only $\sqrt{2/5}N$ transmission peaks exist,
exactly as it is in the case of weakly scattered electromagnetic waves.
However, for stronger disorder where $\xi <L$, only a small
fraction (of order $2\xi/L $) of the states hybridize with the leads. States
which do not hybridize with the leads might have very small transmission,
but nevertheless, they do have a transmission peak. Thus, we expect that $%
N_{T\!R}/L$ will scale with $\xi /L$. Indeed as can be seen in Fig. \ref{Fig4}
, this seems to hold for different values of $L$ and disorder strength $W$.

One can cast the above argument in a more quantitative form. The
overlap of a localized state with the left lead should be proportional to $%
\exp(-b j_0/\xi)$, where $j_0$ is the center of the localized state and $b$
is a numerical constant of order of one depending on the details of the
boundary condition. Averaging over the region $0<j_0<L/2$ for the left lead
and $L/2<j_0<L$ for the right lead, results in: 

\begin{eqnarray}  \label{Eq:OverlapCount}
f = \frac{2}{L} \cdot \sum_{j_0=1}^{L/2} {e^{-b j_0/ \xi}} = \left(\frac{2}{L%
}\right)\frac{1-e^{-b L/ 2 \xi}}{e^{b/\xi}-1} \\
\sim \left(\frac{2 \xi}{b L}\right) \left(1 - e^{- b L/ 2 \xi}\right)  \notag
\end{eqnarray}

Finally, the ratio of the number of transmission peaks to the length of the
wire is obtained by subtracting the fraction of hidden modes times the
probability they overlap with the leads, i.e., 
\begin{equation}
N_{T\!R}/L=1-f\cdot \left( 1-\sqrt{2/5}\right),
\label{Eq:FinalFit}
\end{equation}
which after fitting the parameter $b$ reasonably matches the numerical
results (Fig. \ref{Fig4}).

\begin{figure}[tbp]
\centering
\includegraphics[width=8.0cm,height=!]{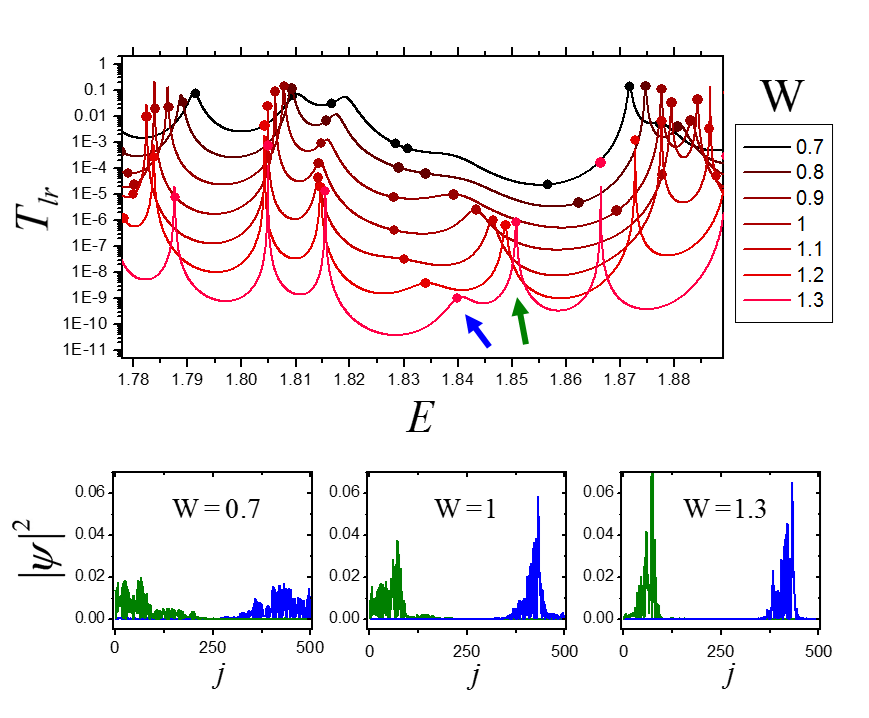}
\caption{\textbf{Upper frame:} The transmission $T_{lr}\left(E\right)$ for a given realization of disorder at different
strengths $W$ from $0.7$ (top black line) to $1.3$ (bottom red line), for a $L=500$ sample with $t_{l/r}=1$. The
eigen-energies of the corresponding isolated wires are marked by circles.
Two modes are hidden at low $W$, and become visible only at higher disorder
level (marked by arrows). \textbf{Lower frame:} The modulus-square of the isolated
eigen-vectors related the two above hidden states. As the disorder
increased, the width of the modes becomes smaller and eventually they
disassociate from the states of the wire.}
\label{Fig5}
\end{figure}

In Fig. \ref{Fig5} we demonstrate the evolution of the transmission spectrum
with increasing strength of disorder. As $W$ grows, the hidden modes
gradually disconnect from the boundaries of the wire and form transmission
resonances, until all of them become ordinary, $N_{T\!R}/L\rightarrow
1$, for large $W$.

It is also interesting to note that the height of the transmission peak is a
non-monotonous function of $W$. While naively one may expect that peaks will
reduce as disorder became stronger, this is correct only on average, and
particular peaks may actually increase when disorder increases.

\begin{figure}[tbp]
\centering
\includegraphics[width=8.5cm,height=!]{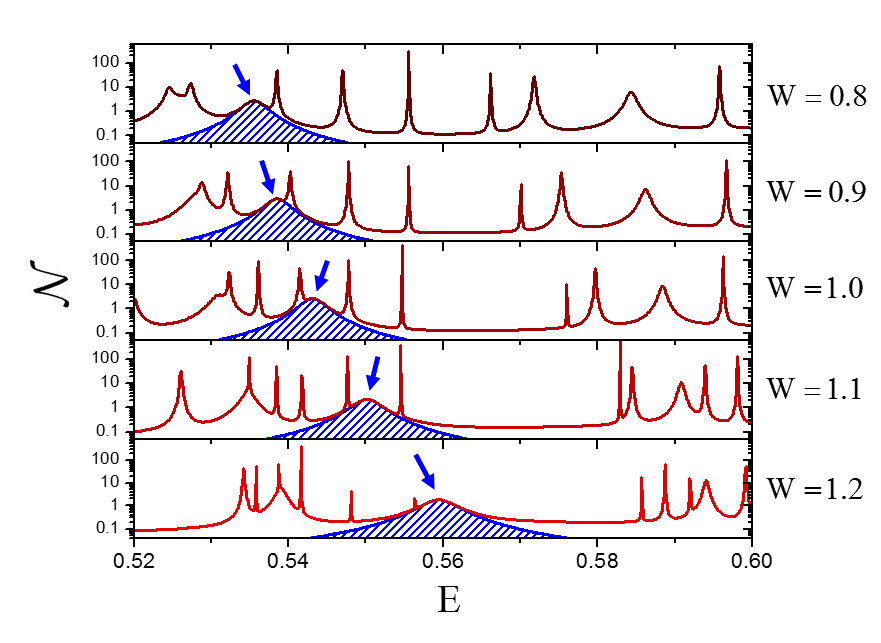}
\caption{Density of states $\mathcal{N}(E)$ of the system depicted in Fig. \ref{Fig5} at different energies, as disorder increases. The ordinary states become
narrower at larger fluctuations, while the hidden mode (marked with blue arrows) widens. Fit to Lorenzian broadening in accordance with Eq. \ref
{Eq:Lorenzian} (blue patterned areas) results in $\gamma_i^{0.8}\!=\!0.00165$, $\gamma_i^{0.9}\!=\!0.00168$, $\gamma_i^{1.0}\!=\!0.00172$, $\gamma_i^{1.1}\!=\!0.00201$ and $\gamma_i^{1.2}\!=\!0.00249$, i.e.  shorter lifetime at the higher disorder level (see text).}
\label{Fig6}
\end{figure}

The spectral broadening of the wire eigenstates (or of the imaginary parts
of the eigenvalues in the Hamiltonian language) is inversely related to
their lifetime. In disordered open systems, as the localization length
becomes shorter (i.e., larger potential fluctuation), one can expect all
modes' lifetimes to increase. This indeed is the case for regular modes, as
seen in Fig. \ref{Fig6}. However, the hidden states again behave
in an unusual way, and remain wide. One can show\cite{Datta1995} that if the
self energy term (Eq. (\ref{Eq:GR_Sigma1})) varies slowly with $E$, the DOS
broadening has a Lorentzian shape: 
\begin{equation}  \label{Eq:Lorenzian}
\mathcal{N}(E) \propto \displaystyle\sum_{i} \frac{\gamma_i}{(E-\tilde{E}%
_i)^2+(\gamma_i)^2}
\end{equation}
where $\gamma_i$ is the imaginary part of the $i_{th}$ eigenvalue, and $%
\tilde{E}_i$ is its real part, modified by the connection to the leads. This
relation allows one to evaluate the lifetime of the $i_{th}$ mode, $\hbar/4\gamma_i$%
, by fitting $\mathcal{N}(E)$ to Eq. (\ref{Eq:Lorenzian}). For the system depicted in Fig. \ref{Fig6}, the life time of the hidden mode at lower disorder
($W=0.8$, $\gamma_i=0.00165$) is longer than at the higher disorder ($W=1.2$%
, $\gamma_i=0.00249$).

\begin{figure}[tbp]
\centering
\includegraphics[width=8.5cm,height=!]{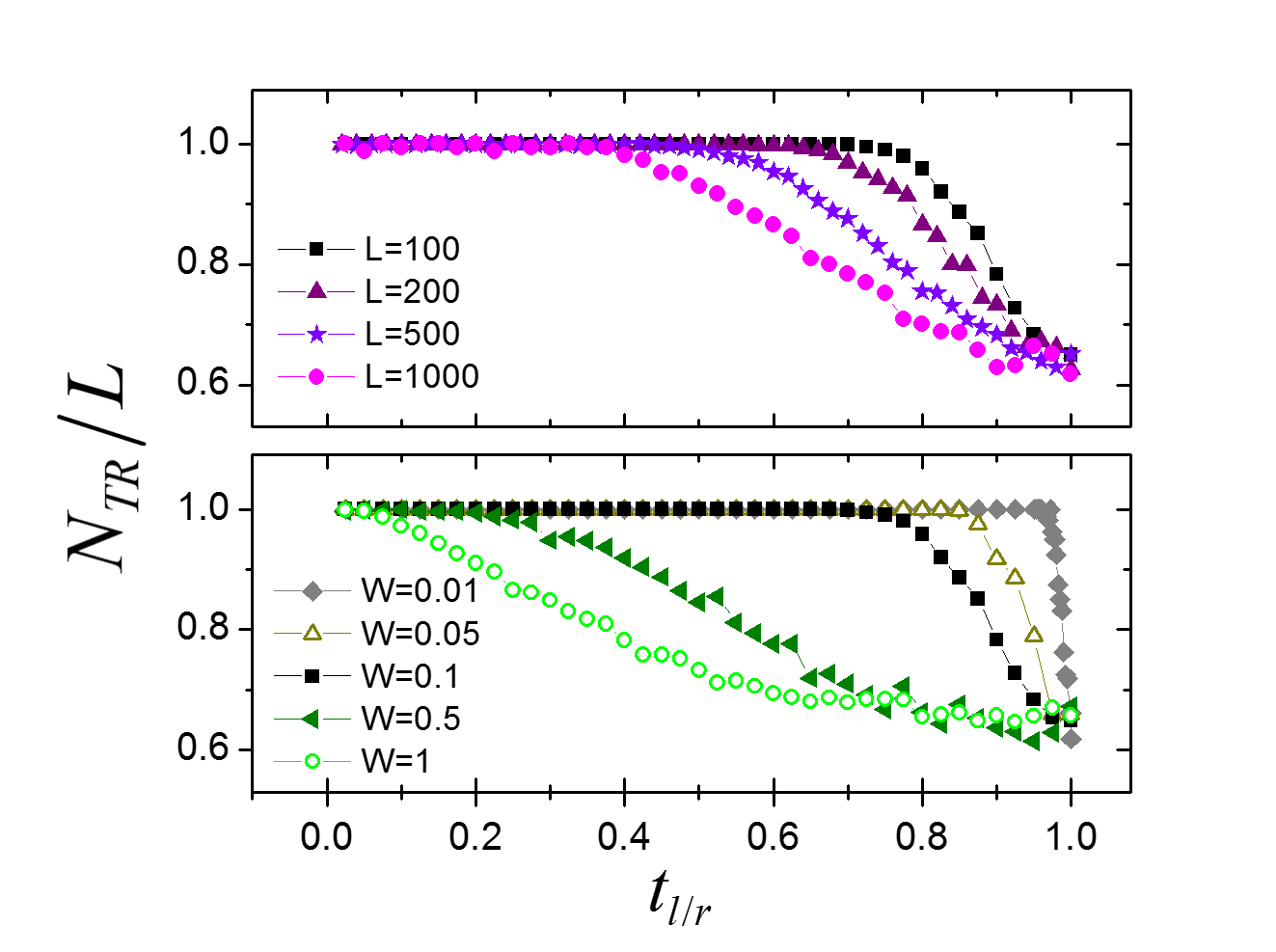}
\caption{The ratio of the number of observed transmission peaks to the
length of the wire $N_{T\!R}/L$ versus lead-system coupling strength, $t_{l/r}$%
. \textbf{Top panel:} $W=0.1$ for different system length $L$. \textbf{Bottom panel:} $L=100$
for different disorder strength $W$. For both cases as $t_{l/r} \rightarrow 0$%
, $N_{T\!R}/L \sim 1$, while for $t_{l/r} \rightarrow 1$ $N_{T\!R}/L \sim 
\protect\sqrt{2/5}$.}
\label{Fig7}
\end{figure}

Since the number of observed transmission resonances depends on both the
disorder and coupling to the environment, the ratio $N_{\mathrm{tr}}/L$ can
be tuned by varying $t_{l/r}$. As this coupling parameter decreases, hidden
modes decouple from the leads and develop peaks in the transmission
spectrum. As can be seen in Fig. \ref{Fig7}, at weak disorder ($W=0.01$) this
transition is sharp: all hidden modes become visible for a very small
change at the vicinity of $t_{l/r} \sim 1$. As the disorder increases (or the system
becomes longer) the coupling amplitude needed to resolve all transmission
resonances becomes smaller and the jump in the ratio $T_{\mathrm{tr}}/L$
broadens. This behavior is counter-intuitive, as one may think that the
enhancement of fluctuations of the potential makes the sample more
``closed'', and therefore will be more easily disconnected from the leads.
In fact, the disorder ties the electronic states strongly to their position in the sample (the edges in the case of hidden modes) and therefore a lower $t_{l/r}$ is required in order to disconnect them. 

The appearance of two time scales when the coupling to the environment
increases and QNMs begin to overlap has been observed in a variety of
regular open physical systems \cite%
{Celardo2009,Celardo2010,Persson2000,Sanchez2011,Morales2012,Aberg2008}; for
a review, see \cite{Auerbach2011} and references therein. This phenomenon is
rather general and is known as the superradiance transition. Its essence is
the following: At weak coupling to the environment the lifetimes of all
states goes down as the coupling increases. As the coupling reaches a
critical value, the states separate into short-lived (superradiant) and
long-lived (trapped) ones, much like the partition of QNMs into ordinary and
hidden modes shown in Fig. \ref{Fig7}.
 However, along with the similarity between the resonance trapping in regular open optical and microwave structures, and between ``hiding'' of some of the resonances in disordered wires there are substantial differences as well.
Indeed, crucial for the superradiance transition are the edge barriers that provide tunable (from very weak and up) coupling of the system with the environment.
Superradiant modes appear in regular systems regardless of disorder, which
just introduces new features (for example, the critical value of the
coupling increases with the degree of disorder \cite{Auerbach2011}) but does not change the essence of the phenomenon.
In the random samples that we consider, finite coupling is implemented by disorder, as the result of the interference of multiply-scattered random fields, even when the system is completely open.
Hidden states appear at the very onset of disorder, when the localization
length is much larger than the size of the samples. When the disorder
increases, the states remain hidden for a wide range of the disorder
strength, and gradually transform into ordinary QNMs as the system reaches
the localized regime.

\begin{figure}[tbp]
\centering
\includegraphics[width=8.0cm,height=!]{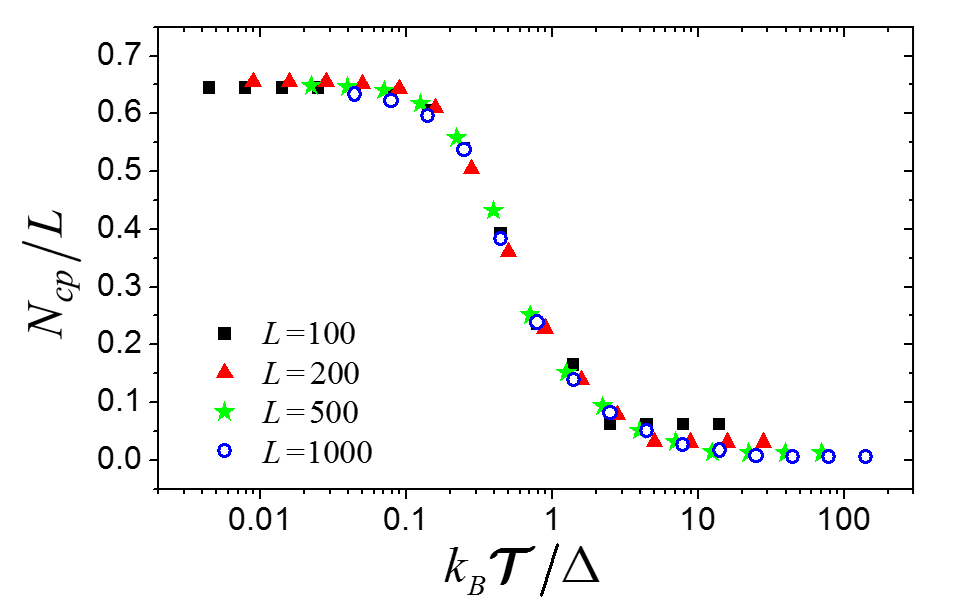}
\caption{The ratio between the number of conductance peaks $N_{cp}$ and the
length $L$ for different values of temperatures $k_B\!\mathcal{T}$ and $L$, for a low bias $V=0.001$ and small disorder $W=0.1$.
The temperature smears some of the conduction peaks and reduces $N_{cp}$ as
it grows. The ratio is scaled by $k_B\!\mathcal{T}/\Delta$ where $\Delta
\sim 4t/L$ is the single level spacing of the disconnected wire. The
transition from zero temperature behavior $N_{cp}/L \sim \protect\sqrt{%
2/5}$ to the high temperature behavior $N_{cp}/L \sim 0$ occurs around $k_B\!\mathcal{T}/\Delta=1$}
\label{Fig8}
\end{figure}

While the transmission is the natural quantity to measure for optical
systems, in electronic systems it is much more commonplace to measure
conductivity. Measuring conductivity is different than measuring
transmission in several aspects. Unlike the ease of generating a single-mode
laser beam, electrons are naturally widely distributed in the energy domain
due to thermal broadening. Therefore, observing the conductance peaks is
possible only if the mean level spacing, $\Delta $, is larger than $k_{B}\!\mathcal{T}$. Thus the ratio of the number of
observable conductance peaks, $N_{cp}$, to the system
length $N_{cp}/L$ falls to zero as $k_{B}\!\mathcal{T}/\Delta \sim 1$.
This crossover for different wire length and temperatures is plotted in
Fig.\ref{Fig8}. The smearing of the conductance peaks for 
$k_B\!\mathcal{T} > \Delta$ is clearly seen.

\subsection{Conclusions}

In this paper, we discussed the effect of disorder on the transmission and
conductivity resonances. We have shown that, similarly to disordered optical
systems, in a 1-D wire with on-site random potential there exists a
ballistic regime, in which a significant amount of eigen-states do not
show a clear peaks in transmission measurements. These `hidden' modes have
extremely broad spectral distributions which, contrary to ordinary Anderson
modes, become even broader (i.e. have shorter life-time) as the disorder
increases. The primary cause of this phenomena is the hybridization with the
states of the attached open leads, which falls off as the localization length 
$\xi $ becomes shorter than the system length $L$, or as the coupling to the
leads is reduced. For weak disorder, the averaged ratio of the number of the
hidden modes to the total number of the electron states in a given energy
interval  deviates only slightly from the constant, $1-\sqrt{\sfrac{2}{5}}$,
as the fluctuations of the potential and/or the length of the wire increase.
This constant coincides with the value analytically calculated in the
single-scattering approximation. The existence of the hidden modes might
substantially affect transport measurements in quantum dots, nanotubes, and
topological insulators, at weak and moderate disorder.

\section{Appendix: Analytical calculation of the ratio $N_{T\!R}/N_{Q\!N\!M}$}

\bigskip Assuming only single scattering process and free electron wave
propagation between scatterers, the transmission probability of an electron
with momentum $k$ in a wire with on-site disorder can be
written as:
\begin{equation}
T\left( k\right) =1-\left\vert r\left( k\right) \right\vert
^{2}=1-\left\vert \sum_{n=1}^{L}r_{n}\cdot e^{i2kan}\right\vert ^{2},
\label{Eq:reflection}
\end{equation}%
where $r_{n}$ is the random reflection amplitude at site $n,$ and $a$ is the
lattice constant. For convenience, we introduce the unit-less length scale so
that $a=1$. Transmission resonances are defined as local maxima of the
transmission coefficient $T\left( k\right) $ so that the resonant values of
the momentum, $k_{n},$\ are the roots of the equation $\frac{dT(k_{n})}{dk}=%
\frac{d\left\vert r\left( k\right) \right\vert ^{2}}{dk}=0$, which can be
presented as 

\begin{equation}
\sum_{n=1}^{N} \sin(2kn)\cdot A_{n}=0.  \label{Eq:TrigonomerticSum}
\end{equation}%
where
\begin{equation*}
A_{n}=\Sigma _{l=1}^{N-n}r_{n+l}r_{l}n+\Sigma _{l=n}^{N}r_{l-n}r_{l}n.
\end{equation*}

Generally speaking, Eq. \ref{Eq:TrigonomerticSum} is a trigonometric polynomial with random
coefficients. The statistics of zeroes of such polynomials have been studied
in \cite{Edelman1995}. Using the results of \cite{Edelman1995} it can be
shown that in a certain interval $\Delta k,$ the ensemble-averaged number of the real roots $N_{\mathrm{root}}$ of the sum in Eq. (\ref{Eq:TrigonomerticSum})
equals to 
\begin{equation}
N_{root}=\frac{2\Delta k}{\pi }\sqrt{\frac{\sum_{l=1}^{N}l^{4}(N-l)}{%
\sum_{l=1}^{N}l^{2}(N-l)}.}  \label{Eq:Roots}
\end{equation}

Calculating the sums in Eq. (\ref{Eq:Roots}) in the limit $N\gg 1$, one gets \cite{Gradshteyn2007}
\begin{equation}
N_{root}\approx \frac{2a\Delta kN}{\pi }\sqrt{\frac{2}{5}}.
\label{Eq:RootSums}
\end{equation}

Since the total number of QNMs in the interval $\Delta k$ is equal to $%
\Delta kLa/\pi ,$ and $N_{T\!R}=N_{root}/2,$ from Eq. (\ref{Eq:RootSums}) it follows
that 
\begin{equation}
\frac{N_{T\!R}}{N_{Q\!N\!M}}=\sqrt{\frac{2}{5}}.  \label{Eq:sqrt25}
\end{equation}

In Fig. \ref{Fig1} it is clearly seen that at the limit of weak disorder ($%
\xi \gg L$) this relation is perfectly followed by the numerical quantum
calculations.


%

\end{document}